\documentclass[times,twocolumn,final]{elsarticle}

\usepackage{medima}
\usepackage{amsmath,amsfonts}
\usepackage{hyperref}
\makeatletter
\def\ps@pprintTitle{%
 \let\@oddhead\@empty
 \let\@evenhead\@empty
 \def\@oddfoot{\reset@font\hfil\thepage\hfil}
 \let\@evenfoot\@oddfoot}
\makeatother

\journal{Medical Image Analysis}

\begin{document}

\verso{Jun Lyu \textit{et~al.}}

\begin{frontmatter}

\title{Rapid Whole Brain Motion-robust Mesoscale In-vivo MR Imaging using Multi-scale Implicit Neural Representation}

\author[1]{Jun \snm{Lyu}\corref{cor1}}
\cortext[cor1]{Corresponding author.\ead{jlyu1@bwh.harvard.edu}}

\author[1]{Lipeng \snm{Ning}}

\author[2]{William \snm{Consagra}}

\author[1]{Qiang \snm{Liu}}

\author[3]{Richard J. \snm{Rushmore}}

\author[4,5,6]{Berkin \snm{Bilgic}}

\author[1]{Yogesh \snm{Rathi}}
\address[1]{Mass General Brigham, Harvard Medical School, MA, United States}

\address[2]{Department of Statistics, University of South Carolina, SC, United States}

\address[3]{Department of Anatomy and Neurobiology, Boston University Chobanian \& Avedisian School of Medicine, MA, United States}

\address[4]{Athinoula A. Martinos Center for Biomedical Imaging, Massachusetts General Hospital, MA, United States}

\address[5]{Department of Radiology, Harvard Medical School, MA, United States}

\address[6]{Harvard/MIT Health Sciences and Technology, MA, United States}

\received{1 May 2013}
\finalform{10 May 2013}
\accepted{13 May 2013}
\availableonline{15 May 2013}
\communicated{S. Sarkar}

\begin{abstract}
High-resolution whole-brain in vivo MR imaging at mesoscale resolutions remains challenging due to long scan durations, motion artifacts, and limited signal-to-noise ratio (SNR).
This study proposes {\underline{Ro}}tating-\underline{v}iew sup\underline{er}-resolution (ROVER)-MRI, an unsupervised framework based on multi-scale implicit neural representations (INR), enabling efficient recovery of fine anatomical details from multi-view thick-slice acquisitions.
ROVER-MRI employs coordinate-based neural networks to implicitly and continuously encode image structures at multiple spatial scales, simultaneously modeling anatomical continuity and correcting inter-view motion through an integrated registration mechanism.
Validation on ex-vivo monkey brain data and multiple in-vivo human datasets demonstrates substantially improved reconstruction performance compared to bi-cubic interpolation and state-of-the-art regularized least-squares super-resolution reconstruction (LS-SRR) with 2-fold reduction in scan time.
Notably, ROVER-MRI achieves an unprecedented whole-brain in-vivo T2-weighted imaging at 180$\mu$m isotropic resolution in only 17 minutes of scan time on a 7T scanner with 22.4\% lower relative error compared to LS-SRR. We also demonstrate improved SNR using ROVER-MRI compared to a time-matched 3D GRE acquisition. Quantitative results on several datasets demonstrate better sharpness of the reconstructed images with ROVER-MRI for different super-resolution factors (5 to 11).
These findings highlight ROVER-MRI’s potential as a rapid, accurate, and motion-resilient mesoscale imaging solution, promising substantial advantages for neuroimaging studies.
\end{abstract}


\end{frontmatter}

\section{Introduction}\label{intro}
Magnetic Resonance Imaging (MRI) is a widely used non-invasive imaging technique for scientific research. The spatial resolution of MRI directly determines the clarity of anatomical details and the accuracy of diagnostic outcomes. However, achieving mesoscale resolution often comes at the cost of reduced signal-to-noise ratio (SNR) and prolonged acquisition times, presenting significant challenges to current research studies~\citep{sui2020slimm}. Balancing resolution, SNR, and acquisition efficiency remains a central issue for MRI technology development~\citep{plenge2012super}.

To accommodate this trade-off, it is common to acquire MRI images with relatively high in-plane resolution and lower through-plane resolution~\citep{zhao2020smore}. In 2D MRI, the through-plane direction is always defined by the slice selection direction, whereas in 3D MRI, the through-plane direction is typically the second phase encoding direction, which usually has the smallest $k$-space sampling width.
While acquisitions with anisotropic voxels may produce diagnostically acceptable images for in-plane directions, they are less suitable for research applications requiring spatially isotropic voxels.

To address these limitations, super-resolution reconstruction (SRR) techniques have been developed.
Super-resolution (SR) techniques are generally categorized into two types: single-image super-resolution (SISR) and multi-image super-resolution (MISR). 
Compared with traditional algorithms~\citep{gholipour2010robust, tourbier2015efficient, manjon2010non,sui2019isotropic,sui2021gradient}, SISR~\citep{bhowmik2017training} methods demonstrate superior performance by learning an end-to-end mapping between pairs of low resolution (LR) and high-resolution (HR) images. 
For example, generative adversarial networks (GANs) have been widely applied in MRI SR tasks. \cite{chen2018efficient} introduced a 3D dense GAN for MRI super-resolution, while \cite{wang2020enhanced} proposed a memory-efficient residual-dense block generator specifically designed for brain MRI super-resolution.
In addition, attention-based models have gained popularity. \cite{zhang2021mr} developed the squeeze-and-excitation reasoning attention network to achieve more accurate MR image super-resolution. \cite{lyu2020mri} introduced an ensemble-based approach to enhance the quality of MR images. \cite{chaudhari2018super} proposed DeepResolve to reconstruct thin-slice high-resolution features from significantly thicker slices.
Recently, \cite{you2022fine} designed a fine perceptive GAN to produce HR MRIs from their LR counterparts, further advancing the field of super-resolution in medical imaging. While these methods can produce excellent results on in-domain examples, they exhibit several notable limitations: (1) large amounts of ground truth data (matched pairs of low and high resolution images) are required for training, and (2) limited generalization capability to out-of-distribution tasks such as super-resolution from different contrasts, and MR images with tumors or lesions or other disease-related biological features not present in training data.  

In contrast, MISR exploits information from multiple complementary views or different contrasts from the same subject~\citep{wu2021irem, rousseau2010non}. 
For example, \cite{greenspan2002mri} introduced iterative back-projection algorithms to improve slice-direction resolution using slightly shifted multislice data. \cite{scherrer2012super} enhanced spatial resolution of diffusion MRI through orthogonal anisotropic acquisitions combined with distortion correction and maximum a posteriori reconstruction. \cite{vis2021accuracy} utilized multi-orientation, thick-slice acquisitions for submillimeter quantitative MRI with optimized inter-slice reconstruction. More recently, \cite{dong2025romer} proposed Romer-EPTI, a rotating-view, motion-robust superresolution technique that integrates echo-planar time-resolved imaging (EPTI), achieving mesoscale diffusion MRI with minimal distortion and blurring. However, the reconstruction technique used in this work is the standard least-squares fitting as described in \citep{vis2021accuracy}.
\cite{zeng2018simultaneous} introduced a two-stage deep convolutional neural network capable of performing both single- and multi-contrast SR reconstructions. \cite{lyu2020multi} proposed a progressive network designed for high up-sampling multi-contrast MRI, which learns shared features in a joint representation space. \cite{feng2021multi} further advanced the field with  multi-stage integration networks and separable attention modules~\citep{feng2024exploring} to restore target-contrast images by leveraging auxiliary contrast images through cross-contrast feature exploration.
Moreover, transformer-enabled frameworks~\citep{li2023multi, lyu2023multicontrast, li2022transformer, li2022wavtrans, li2024rethinking, li2023dudoinet} have been used to enhance joint feature space learning, providing a significant enhancement in multi-resolution SR performance.
However, all these SRR methods typically require extensive high-resolution ground truth datasets for training, which are scarce or not available at mesoscale ($<0.5mm$) resolution. Further, these methods generalize poorly to new types of data~\citep{willemink2022toward} or to the presence of biological abnormalities such as tumors or lesions. 

Another alternative for mesoscale imaging is to use ultra-high magnetic field strength such as the 11.7T scanner,, which reported 0.55mm isotropic resolution for T2w imaging~\citep{boulant2024vivo} or a specialized head-only scanner such as the Impulse 7T scanner~\citep{feinberg2023next} which demonstrated a resolution of 0.45mm isotropic voxel size. However, such scanners are one-of-a-kind scanners not widely available.

Implicit Neural Representations (INRs) have emerged as a powerful approach for learning continuous signal representations from discrete samples in an unsupervised manner without requiring any ground truth for training the network~\citep{sitzmann2020implicit, dwedari2024estimating, consagra2024neural, yu2025bilevel}. Unlike traditional methods that store signal values discretely on coordinate grids, INRs represent the signal as a continuous function using trainable neural networks, particularly multi-layer perceptrons (MLPs)~\citep{mildenhall2021nerf}. By approximating the complex relationships between coordinates and their corresponding signal values, INRs produce continuous signal representations. INRs fall in the class of self-supervised learning techniques that utilize inherent patterns within the data itself, removing reliance on explicitly paired datasets and thereby generalizing to diverse tasks. The flexibility and effectiveness of INRs have led to their application in various medical imaging tasks \citep{ye2023super, mcginnis2023single}, including supervised 3D super-resolution~\citep{wu2022arbitrary}, tumor progression assessment from sparsely sampled images~\citep{shen2022nerp}, and isotropic reconstruction from a single anisotropic image~\citep{zhang2023self}. However, none of these techniques address the problem of mesoscale super-resolution from multi-view thick slice data without a-priori training.

\begin{figure*}[t]
\centerline{\includegraphics[width=\textwidth, angle=0]{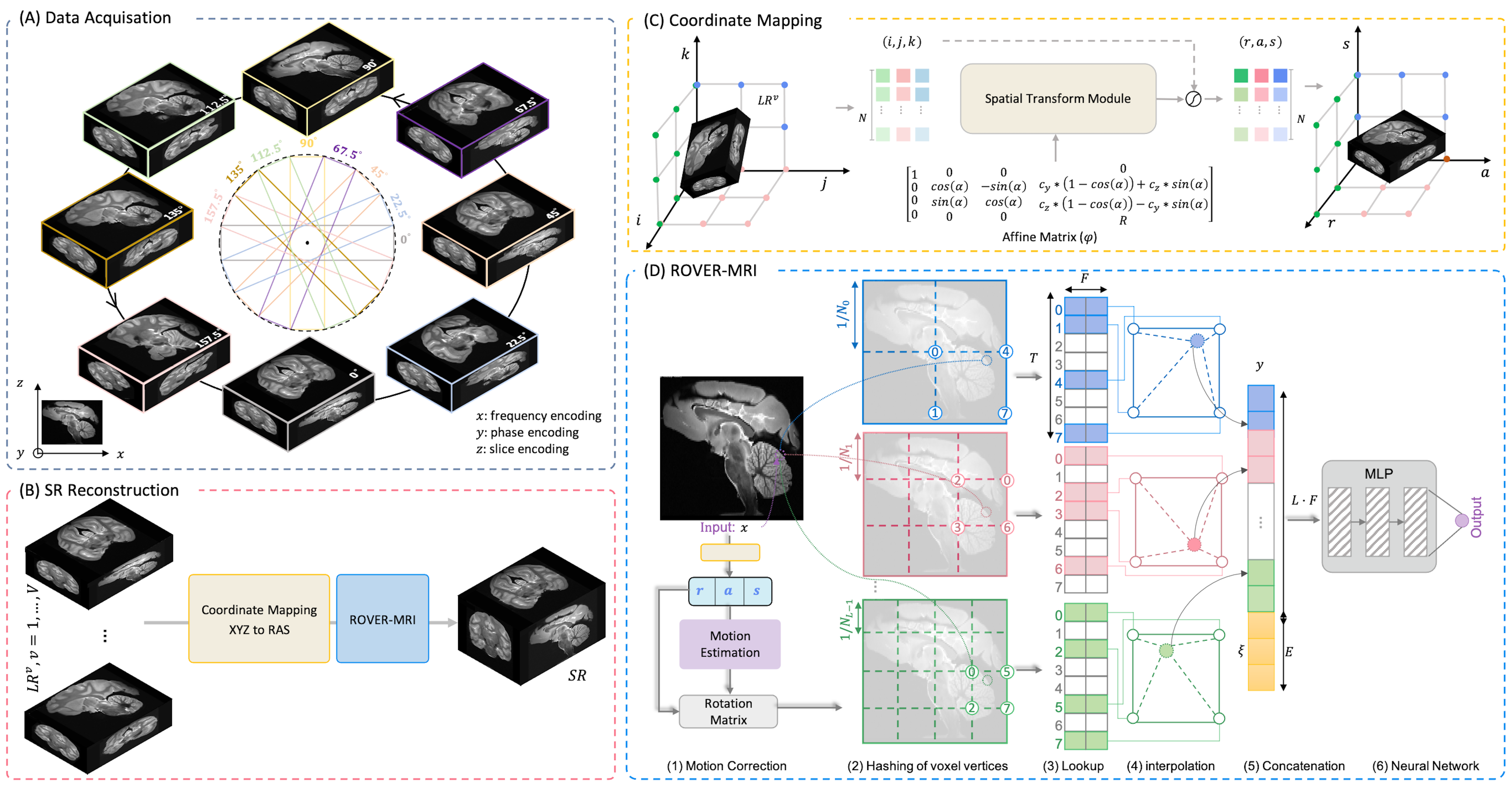}}
\caption{Overview of the ROVER-MRI framework. \textbf{(A) Data Acquisition}: Eight low-resolution images are acquired with varying slice orientations to ensure comprehensive spatial coverage. \textbf{(B) SR Reconstruction}: A neural network-based implicit representation maps RAS spatial coordinates to pixel values, enabling high-quality super-resolution reconstruction. \textbf{(C) Coordinate Mapping}: Matrix coordinates from low-resolution images are preprocessed to extract precise image values for reconstruction. \textbf{(D) ROVER-MRI}: Multi-resolution hash encoding integrates spatial mapping, voxel hashing, feature retrieval, and auxiliary input concatenation to achieve enhanced reconstruction performance.}\label{fig1}
\end{figure*}

Existing periodic activation function based INRs, termed as SIREN-based INR's~\citep{sitzmann2020implicit}, process point-wise coordinates using a single-head MLP. However, they often struggle to fully capture both local details and global features in large-scale signals, as training the extremely large MLPs required for such tasks is challenging~\citep{xie2022neural}, thereby limiting the reconstruction quality. To overcome these challenges, a multi-scale hashing encoding-based INR method has been proposed ~\citep{muller2022instant, dwedari2024estimating}. By mapping spatial coordinates into compact multi-scale hash tables, this approach significantly enhances computational and storage efficiency while effectively capturing multi-resolution information, thus improving the model’s representational capacity. We propose to use this method in our work to reconstruct multi-view MRI data. 

 \section{Our contributions}
In this work, we propose several novel contributions to enable rapid mesoscale MRI. 
\begin{enumerate}

\item We develop a multi-scale INR method that learns a continuous image reconstruction from multi-view (rotated) thick slice MR images. The continuous model is then sampled to produce a single high isotropic resolution MR image volume. 
We note that, this is the first time INRs have been tailored to represent such high resolution data (each image dimension with more than 1000 voxels) in the medical imaging context. 

\item We validate the performance of our method on $125 \mu m$ ex-vivo monkey data, where ground truth data is available. We also quantitatively compare our method with the current state-of-the-art method for multi-view reconstruction, termed least-squares super-resolution reconstruction (LS-SRR)~\citep{vis2021accuracy, dong2024romer}.

\item We demonstrate excellent performance of our method from sparse set of views, thereby allowing to reduce scan time by a factor of 2. We demonstrate the performance of our method on several challenging in-vivo human datasets, but notably, we demonstrate the ability to reconstruct the highest resolution to-date of whole brain in-vivo human T2w image at $180 \mu m$ isotropic resolution in 17 minutes. 

\item We incorporate motion correction in our reconstruction framework and demonstrate the efficacy of the method in recovering details in simulated as well as real head motion (inside the scanner) settings. 

\item Finally, we show that ROVER-MRI can be used to achieve much higher spatial resolution and SNR than 3D acquisitions in scan time-matched gradient-echo (GRE) scans.

\end{enumerate}


\section{METHODS}
\subsection{Acquisition Strategy}
To achieve super-resolution reconstruction, multiple low-resolution datasets are combined to achieve high isotropic resolution. In this study, as depicted in Figure~\ref{fig1}(A), we acquired thick-slice images (but high in-plane resolution) by systematically rotating the acquisition around the phase-encoding axis, following the multi-stack rotational acquisition method introduced by \cite{shilling2008super}. The theoretical minimum number of rotations ($N_R$) required to reconstruct isotropic voxel resolution is determined based on the geometric framework for super-resolution reconstruction developed by \cite{plenge2012super}, as expressed in the following equation:
\begin{equation}
\label{eq1}
N_R \geq \frac{\pi}{2} \cdot \alpha,
\end{equation}
where $\alpha$ represents the ratio of the longer to shorter dimensions of a voxel. Additionally, as noted by \cite{edelstein1986intrinsic}, the SNR in multi-slice imaging is strongly influenced by voxel volume, with larger voxels typically providing enhanced SNR.

The generation of a low-resolution image ($\mathbf{LR}^v$) from a high-resolution image ($\mathbf{y}$) can be described mathematically as follows: 
\begin{equation}
\label{eq2}
\mathbf{LR}^v=\mathbf{H}_v \mathbf{y}+\mathbf{n}_v,
\end{equation}
where $v$ denotes a specific low-resolution sample, while $\mathbf{n}_v$ is random Gaussian vector modeling the measurement noise. The matrix $\mathbf{H}_v$ encapsulates the cumulative effects of geometric transformations, resolution downsampling, and blurring applied to the high-resolution input. This matrix is influenced by factors such as image dimensions, anisotropic voxel sizes, and the slice profile, which is modeled in this study using a rectangular function.
\subsection{Coordinate Mapping}
Figure~\ref{fig1}(B) illustrates the coordinate mapping process used in super-resolution reconstruction, showcasing the transformation of low-resolution voxel coordinates $(i,j,k)$ to a unified high-resolution real-world coordinate system $(r,a,s)$.

First, the low-resolution input image ($LR^v$ ) is represented in the original coordinate system, denoted by $(i,j,k)$, with distinct voxel locations visualized as green, pink, and blue points. This representation encapsulates the anisotropic resolution of the original acquisition. 
Then, to map the low-resolution data into a common high-resolution space, an affine transformation is applied. The transformation matrix ($\varphi$) includes both rotational and translational components. Rotation is parameterized by an angle $\alpha$, while translation accounts for slice-specific offsets, such as $c_y$ and 
$c_z$. The last element in the fourth column, $R$ represents the scale in space (commonly $R=1$, indicating no scaling).
The affine matrix is defined as:
\begin{equation}
\varphi=\left[\begin{array}{cccc}
1 & 0 & 0 & 0 \\
0 & \cos (\alpha) & -\sin (\alpha) & c_y(1-\cos (\alpha))+c_z \sin (\alpha) \\
0 & \sin (\alpha) & \cos (\alpha) & c_z(1-\cos (\alpha))-c_y \sin (\alpha)\\
0 & 0 & 0 & R
\end{array}\right],
\label{eq3}
\end{equation}
enabling the systematic alignment of slices in the reconstructed coordinate system. The affine-transformed coordinates are passed through a spatial transform module, which interpolates the voxel intensities from the low-resolution grid $(i,j,k)$ to the high-resolution coordinate space $(r,a,s)$. This module ensures spatial consistency and accuracy during the reconstruction process.
Finally, a motion estimation block was introduced, where rotations between different acquisition views and the reference view (first acquired view) are estimated using affine registration provided by FSL~\citep{jenkinson2012fsl}. Subsequently, the high-resolution voxel grid in the $(r,a,s)$ coordinate space is transformed using these estimated rotation matrices.

\subsection{Image reconstruction using ROVER-MRI}
Figure~\ref{fig1}(D) illustrates our SRR technique using coordinate mapping and ROVER-MRI. The input to the network are the $(r,a,s)$ spatial coordinates in the image field, while the output is the corresponding pixel value.  

\subsubsection{Implicit Neural Representation}
INRs model signals, such as images or volumetric data, as continuous and differentiable functions parameterized by neural networks. INRs take spatial coordinates $\mathbf{x}_i$ as input and output the corresponding signal value $y_i$, formally expressed as:
\begin{equation}
\mathcal{M}(\boldsymbol{\theta}) : \mathbb{R}^\mathrm{dim} \to \mathbb{R}, \quad \mathbf{x}_i \mapsto \mathcal{M}(\boldsymbol{\theta}; \mathbf{x}_i) = y_i,
\end{equation}
where $\mathcal{M}$ represents a neural network parameterized by $\theta$, and $\operatorname{dim}$ denotes the dimensionality of the spatial domain.

This framework allows signals to be represented efficiently at any desired resolution by querying their values at continuous spatial coordinates. Such a representation is particularly advantageous in tasks like encoding image data~\citep{sitzmann2019deepvoxels}, modeling 3D shapes~\citep{gao2022get3d}, and performing neural rendering~\citep{wysocki2024ultra}. Early methods that utilized sinusoidal encodings within a single global MLP-based representation often struggle to capture high-frequency details in large-scale signals. These issues arise from prohibitively long training times and diminishing representational capacity of the very large MLPs necessary for such a global parameterization~\citep{xie2022neural}. To address these challenges, recent innovations have introduced advanced encoding techniques~\citep{hamming1952mathematical}, improved activation functions~\citep{sitzmann2020implicit,saragadam2023wire}, and optimized network structures~\citep{kazerouni2024incode,zhu2024disorder}, all of which significantly enhance the representational power of INRs.

One notable advancement is the use of position-aware architectures, often paired with multi-resolution encoding techniques like hash grids, to capture fine details in signals without incurring high memory costs. By leveraging these developments, INRs have proven to be highly effective at modeling complex signal structures while remaining computationally efficient. These advancements have positioned INRs as a cornerstone of modern machine learning, enabling precise and continuous signal representations across a variety of applications.

\subsubsection{Multi-scale Hash Grid Encoding}
The multi-resolution hash grid representation is a pivotal technique designed to enhance the efficiency of INRs while maintaining high spatial fidelity~\citep{muller2022instant}. This approach employs a multi-resolution hash grid encoder to map input coordinates ${x} \in[0,1]^d$, where $d$ is the spatial dimension, into high-dimensional feature vectors $\mathbf{h} \in \mathbb{R}^{L \cdot F}$. These feature vectors of dimension $F$ are parameterized across $L$ resolution levels, with each level storing its features in hash table $H$. This hierarchical representation captures features across varying spatial scales, enabling efficient encoding of both global and local details.

To extract features, the input domain is divided into 
$L$ levels of grids, where each grid level $l$ has a resolution $N_l$ defined as:
\begin{equation}
N_l=\left\lfloor N_{\min } \cdot b^{l-1}\right\rfloor, \quad b=\exp \left(\frac{\ln N_{\max }-\ln N_{\min }}{L-1}\right).
\end{equation}
The scaling factor $b$ ensures that grid resolutions increase exponentially from the coarsest resolution 
$N_{min}$  to the finest $N_{max}$. At the finest levels, the vertex count grows significantly, but memory overhead is controlled by using hash tables for feature storage. The feature lookup is performed via a hash function:
\begin{equation}
h(\mathbf{x})=\left(\bigoplus_{i=1}^d x_i \pi_i\right) \bmod T,
\end{equation}
where $\pi_i$ are unique, large prime numbers, $T$ denotes the hash table size, and $\oplus$ is the bitwise XOR operation. This hashing mechanism ensures efficient memory utilization by assigning unique indices to grid vertices while minimizing hash collisions. Due to the multi-resolution structure, the likelihood of collisions across all levels is negligible, even as the resolution increases.
Finally, the encoded features are passed to a fully connected neural network 
where the network resolves the target function or vector field. This combination of multi-resolution hash grids and compact neural networks supports high spatial resolution while minimizing memory usage. This architecture is particularly effective for applications requiring fine spatial details in large-scale signals.

For a given input point, features are interpolated within each grid level by calculating weights 
$\mathbf{w}_l=\mathbf{x}_l-\left\lfloor\mathbf{x}_l\right\rfloor$, where $\left\lfloor \right\rfloor$ is the floor operator. These weights enable smooth blending of feature values between grid vertices. The interpolated feature vectors from all $L$ levels are then concatenated and optionally combined with auxiliary inputs $\xi$, which represent the spatial coordinates themselves, forming the encoded representation $\mathbf{h}=\operatorname{enc}(\mathbf{x} ; \xi)$.

As illustrated in Figure~\ref{fig1}(D), the entire process of reconstructing the final isotropic resolution image can be summarized in the following steps: (1) we map the input coordinates to a common RAS coordinate system through an affine transformation. (2) We then hash the 3D image coordinates into grid cells, efficiently assigning unique codes to each voxel for optimized memory and computation. (3) These hash codes are used to look up the corresponding feature embeddings from a hash table, greatly reducing processing time. (4) Next, we apply linear interpolation to these embeddings to produce smooth, high-resolution transitions across the grid. (5) Features from multiple resolution levels are concatenated, enabling the model to capture fine details and broader structures. (6) Finally, a MLP maps the combined features to predict the super-resolved image.

\subsection{Loss Functions}
To achieve high-quality image reconstruction, the proposed method incorporates a loss function specifically designed for super-resolution tasks. It includes two main components: a reconstruction loss and total variation (TV) regularization. The reconstruction loss enforces consistency between the predicted output and the LR acquired data, ensuring accurate data representation. Meanwhile, TV regularization acts as a prior, promoting smoothness and reducing noise or artifacts in the reconstructed images. By combining these two components, the framework effectively balances precise alignment with the input data and the generation of visually appealing and coherent outputs.

\subsubsection{Reconstruction Loss}
The reconstruction loss is formulated using the Mean Squared Error (MSE), adapted for super-resolution tasks. The high-resolution predictions are averaged along the super-resolution direction to generate a downsampled version that matches the resolution of the low-resolution ground truth $LR^v$. The loss is defined as:

\begin{equation}
\mathcal{L}_{\mathrm{MSE}}=\frac{1}{N} \sum_{i=1}^N\left\|L R_i^v-\hat{y}_i^{\mathrm{avg}}\right\|_{\textcolor{black}{2}}^2
\end{equation}%
where $LR^v$ denotes the low-resolution ground truth, $\hat{y}_i^{\mathrm{avg}}$ is obtained by passing the estimated high-resolution predictions through the $\mathbf{H}_v$ operator, and $N$ is the total number of data points. By penalizing the squared differences between the averaged predictions and the low-resolution ground truth, the MSE loss ensures that the model outputs are consistent with the observed data. This loss is crucial for guiding the network to accurately capture the underlying structure of the input and is particularly well-suited for image super-resolution and restoration tasks.

\subsubsection{TV Regularization}
Total Variation (TV) regularization is a gradient-based prior designed to improve the structural consistency of reconstructed images. By penalizing gradient variations in the predicted output, it promotes smoothness in regions of uniform intensity while preserving sharp edges. \textcolor{black}{In our implementation, the image gradient is approximated using finite differences in the discrete domain. Specifically, we compute first-order differences along spatial dimensions to estimate the gradient magnitude.} The TV loss is formulated as:
\begin{equation}
\mathcal{L}_{\mathrm{TV}}=\frac{1}{M} \sum_{j=1}^M\left\|\nabla \hat{y}_j\right\|_{\textcolor{black}{1}},
\end{equation}
where $\nabla \hat{y}_j$ represents the gradient of the predicted output, and $M$ is the total number of samples. This regularization effectively reduces noise and suppresses unwanted high-frequency artifacts, which are often present in inverse problems like image reconstruction.

To incorporate TV regularization into the overall optimization process, a weighting parameter $\lambda_c$ is introduced to control its contribution. The combined loss function is expressed as:
\begin{equation}
\mathcal{L}=\mathcal{L}_{\mathrm{MSE}}+\lambda_c \cdot \mathcal{L}_{\mathrm{TV}}.
\label{eq9}
\end{equation}
Here, $\lambda_c$ balances the trade-off between the data fidelity term ($\mathcal{L}_{\mathrm{MSE}}$) and the smoothness prior ($\mathcal{L}_{\mathrm{TV}}$). Adjusting $\lambda_c$ allows the framework to prioritize either reconstruction accuracy or structural regularity, depending on the specific demands of the task. This flexibility ensures that the approach can adapt to various scenarios while maintaining high-quality outputs. 
\textcolor{black}{The entire code for our ROVER-MRI framework is available at \href{https://github.com/rebeccalyu666/ROVER_MRI}{https://github.com/rebeccalyu666/ROVER-MRI}.}
\begin{figure*}[t]
\centerline{\includegraphics[width=0.8\textwidth, angle=0]{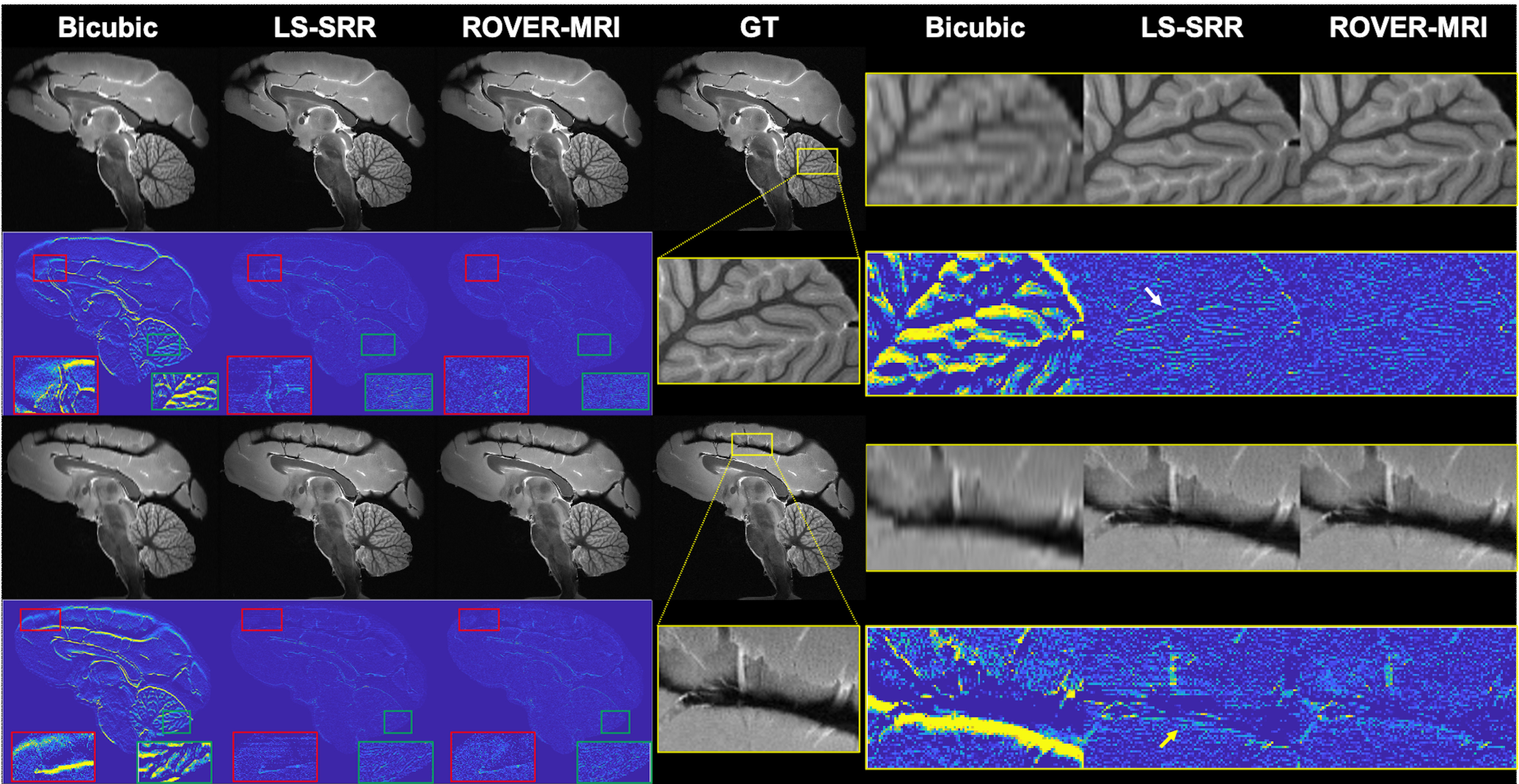}}
\caption{Reconstruction results on simulated LR data using Bicubic interpolation, LS-SRR, and ROVER-MRI. The first and third rows show reconstructed sagittal MRIs, while the second and fourth rows present the corresponding error maps calculated against the GT. The red and green boxes highlight zoomed-in regions, which allow a closer inspection of the reconstruction quality. 
}\label{fig2}
\end{figure*}
%
\subsection{EXPERIMENTS}
\subsubsection{Datasets}
We performed extensive experiments to evaluate ROVER-MRI using the following four datasets. The experiments in this study were conducted following approval from the local Institutional Review Boards (IRBs). Informed consent was provided before the in-vivo human scan.
\begin{itemize}
\item \textbf{Ex-vivo monkey MRI Data} 
We acquired high-resolution T2-weighted ex-vivo images of a monkey brain on a 9.4T Bruker scanner (PharmaScan; Bruker BioSpin, Ettlingen, Germany) with a spatial resolution of 0.125$\times$0.125$\times$0.125 mm$^{3}$ using a 3D RARE sequence. The imaging parameters were: TR = 1.0 s, TE = 0.02 ms, Flip Angle(FA) = 180 degree. This dataset served as the ground truth for all validation purposes with low-resolution data (0.125$\times$0.125$\times$0.625 mm$^{3}$) from different views generated through simulation (downsampling) for each of the 8 views separated by 22.5$^\circ$ to satisfy the Nyquist criteria for a super-resolution factor of 5 in the slice dimension. 
%
Additionally, inter-view motion was simulated by introducing rotations of 5°, 7°, and 10° to three of the eight views to evaluate the robustness of motion correction.

\item \textbf{Human in-vivo 7T MRI Data}
For Participant 1 (P1), T2-weighted SPACE images were acquired using a 7T Siemens Terra scanner (Siemens Healthineers, Erlangen, Germany, with a single transmission 32-channel head coil) at 15 distinct rotation angles, separated by 12$^\circ$. Each of the rotated views were acquired at a resolution of 0.18$\times$0.18$\times$2.00 mm$^{3}$, with a total scan time of approximately 34 minutes and a super-resolution factor of 11 in the slice dimension.

For Participant 2 (P2), 2D-GRE images were acquired (7T Siemens Terra scanner) at 9 distinct rotation angles, separated by 40$^\circ$. The spatial resolution for these images was 0.4519$\times$0.4519$\times$2.70 mm$^{3}$ with the slice dimension being 6 times the in-plane resolution. The imaging parameters were: TR = 1.45 s, TE = 4.67 ms, FA = 77 degree, acceleration factor PE = 3, number of slices = 61. The total scan time was ~36 minutes. Further, to qualitatively compare ROVER-MRI reconstruction with a full 3D acquisition, we acquired another scan (3D GRE) with a spatial resolution of 0.45mm isotropic voxels. The imaging parameters were: TR = 24.0 ms, TE = 4.67 ms, FA = 13 degree, acceleration factor PE = 3, slices per slab = 288. The scan time for this scan was 18:38 minutes.

For Participant 3 (P3), 2D-GRE images were acquired (7T Siemens Terra scanner) at 8 distinct rotation angles, separated by 22.5$^\circ$ with the participant instructed to move their head randomly during the scan to mimic realistic head motion scenarios. The spatial resolution for these images was 0.18$\times$0.18$\times$2.00 mm$^{3}$ with the slice dimension being 11 times the in-plane resolution. The imaging parameters were: TR = 10.3 s, TE = 44 ms, FA = 120 degree, percent sampling: 75\% , number of slices = 74. The total scan time was about 27 minutes. Further, to qualitatively compare ROVER-MRI reconstruction with a full 3D acquisition, we acquired another scan (3D GRE) with a spatial resolution of 0.2$\times$0.2$\times$0.5 mm$^{3}$. This was the highest possible resolution that could be acquired on the scanner. The imaging parameters were: TR = 10.3 s, TE = 51 ms, FA = 120 degree, percent sampling: 75\%, slices per slab = 256. The scan time for this scan was about 23 minutes.

\item \textbf{Human in-vivo 3T T2w MRI Data}
On a 3T Siemens Prisma scanner (32 channel head coil), we acquired T2-weighted images at 8 different rotation angles, spaced 22.5$^\circ$ apart. The images were captured with a spatial resolution of 0.42$\times$0.42$\times$2.12 mm$^{3}$ to enable a super-resolution reconstruction by a factor of 5 in the slice dimension. The imaging parameters were: TR = 2.5 s, TE = 0.202 ms, FA = 120 degree. 

To acquire low-resolution datasets, slices were rotated around a stationary axis oriented along the anterior-posterior direction. The field-of-view (FOV) was strategically designed to include the entire object with sufficient margins, ensuring precise coverage for SRR. The minimum number of rotation angles required for each dataset was computed using Eq.\ref{eq1}. 
In all experiments, high-resolution images were reconstructed from the low-resolution data, following the framework described in section 3.3.2 (Figure~\ref{fig1}(D)).
\end{itemize}

\subsubsection{Implementation}
All RAS coordinates and MR image intensities were normalized to [0,1]. 
In all experiments, we use an MLP with two hidden layers that have a width
of 192 neurons, SineLayer~\citep{sitzmann2020implicit} on their hidden layers, and a linear output layer. The hash encoding parameters and $\lambda_c$ in the loss function were set as given in Table~\ref{tab1}.
The training utilized the Adam optimizer with a learning rate of $10^{-4}$ over 10000$\sim$28000 iterations on a single A100 GPU. For comparison, we benchmarked our approach against two SRR methods: bicubic and least-squares SRR (LS-SRR)~\citep{vis2021accuracy}. 
Both visual comparison and quantitative evaluation were used for performance evaluation. For a quantitative evaluation, the relative error (RE) was calculated as follows:
\begin{equation}
\text { Relative Error }=\frac{1}{V} \sum_{v=1}^V\frac{| { y^{v} }- { LR }^{v} |_{\textcolor{black}{1}}}{ { |LR^{v}|_{\textcolor{black}{1}}}}
\end{equation}
where $LR^v$ and $y^v$ represent the low-resolution thick-slice acquired data and the back-projection from the high-resolution reconstructed image respectively.
%
To quantitatively compare image sharpness across reconstruction methods, a common region of interest (ROI) was manually selected. Image sharpness was measured using the Laplacian variance method~\citep{pech2000diatom}, where a Laplacian filter ($\alpha=0.2$) was applied to the ROI. Specifically, sharpness was calculated as the variance of the Laplacian-filtered intensities:
\begin{equation}
\text {Sharpness}=\text{Var[$L_{x,y}$]}=\frac{1}{MN} \sum_{x=1}^M\sum_{y=1}^N(L(x,y)-\mu)^2
\end{equation}
where $L_{x,y}$ denotes the filtered image, $M$$\times$$N$ represents the ROI size, and $\mu$ is the mean intensity of $L_{x,y}$. Higher Laplacian variance indicates superior image sharpness, reflecting stronger edges and richer high-frequency details


\section{RESULTS}
\label{sec:RESULTS}
\subsection{Experiment 1: Validation on ex-vivo monkey MRI with known ground truth}
%


%
We first validated our proposed method on ex-vivo monkey T2-weighted MRI data, where ground truth is available, providing a reliable basis for quantitative and qualitative comparisons. Figure~\ref{fig2} illustrates the reconstruction results using bicubic, LS-SRR and our proposed ROVER-MRI method, highlighting the differences in performance across methods. 
The bicubic interpolation method exhibits pronounced blurring of fine cerebellar structures and brain tissue boundaries, as evident in the magnified regions (yellow boxes). The LS-SRR method improves sharpness but introduces noticeable streaking artifacts, compromising reconstruction quality. Additionally, the error maps show that the LS-SRR method struggles to preserve fine structural boundaries (see red and green boxes in Figure~\ref{fig2}). While LS-SRR performs better than bicubic interpolation, these artifacts and errors limit its effectiveness in preserving anatomical accuracy.

In contrast, ROVER-MRI demonstrates superior performance by minimizing residual errors and effectively restoring fine details with high fidelity. 
%
Quantitative sharpness evaluation, measured by Laplacian variance within the two yellow-boxed regions shows higher sharpness using ROVER-MRI method (0.0105 and 0.0123) compared to LS-SRR (0.0089 and 0.0104), underscoring its superior capability in preserving fine anatomical details.

%
Table~\ref{tab2} summarizes the Relative Error (RE) comparisons for different MRI reconstruction methods (LS-SRR and ROVER-MRI) across various imaging scenarios. The results demonstrate that ROVER-MRI consistently outperforms SRR in most conditions, achieving lower RE by approximately 22.4\% compared to LS-SRR. 
The clear preservation of structural integrity, particularly in the cerebellum, underscores the superior performance of our approach. These results highlight the robustness of our method in maintaining anatomical precision, which is crucial for downstream analyses.

\subsection{Experiment 2: Validation with fewer views and low SNR data}
\begin{figure*}[t]
\centerline{\includegraphics[width=0.8\textwidth, angle=0]{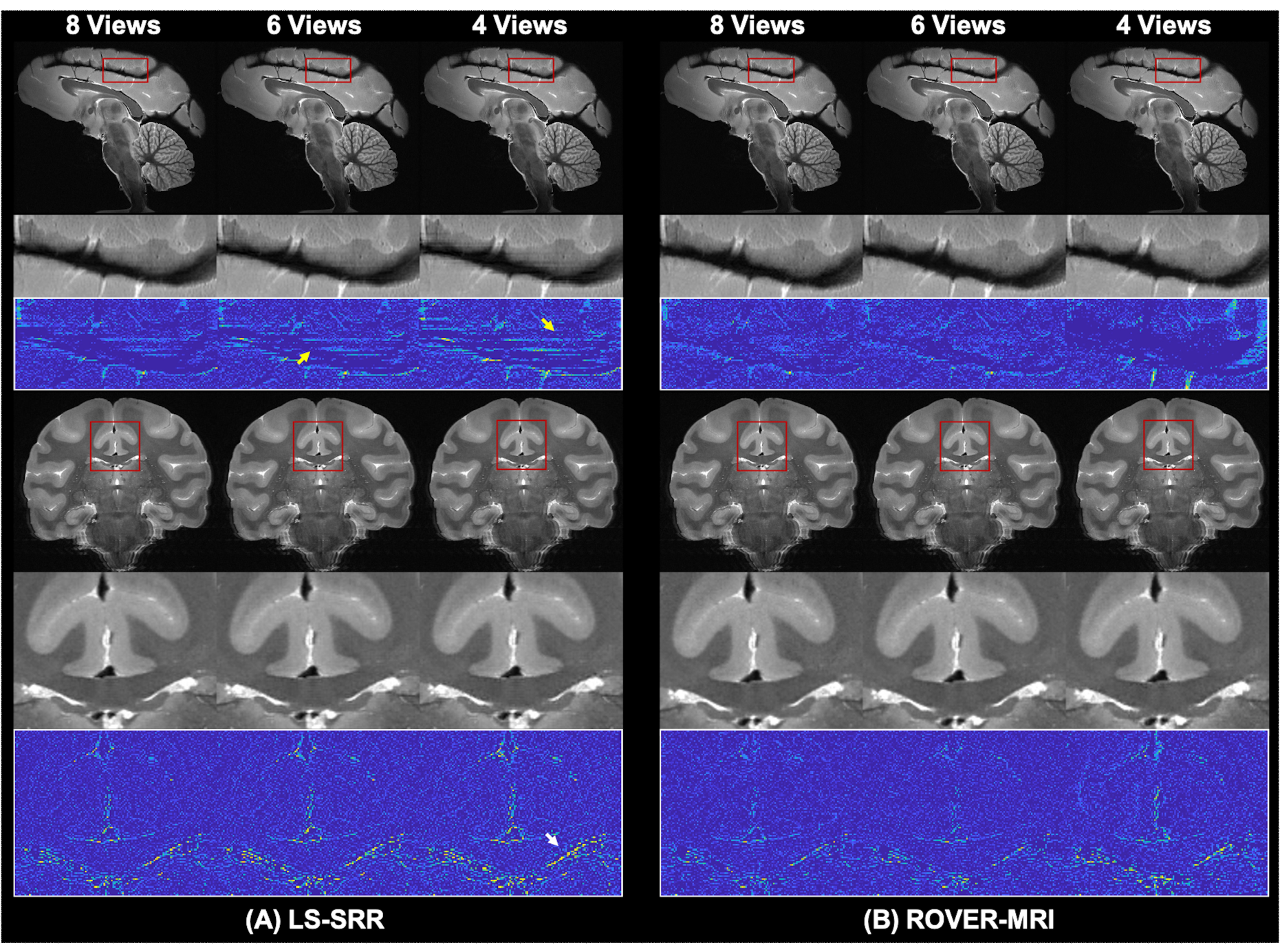}}
\caption{Reconstruction results of our method and LS-SRR using fewer views. (A) LS-SRR reconstruction results, showing increased error as the view count decreases. (B) ROVER-MRI reconstruction results, demonstrating high-quality reconstruction even with fewer views. Rows 1 and 4 display typical reconstruction results, while Rows 2 and 5 show enlarged views of the regions within the red boxes. Rows 3 and 6 show the error maps for these regions, highlighting the superior reconstruction accuracy of ROVER-MRI compared to LS-SRR.}\label{fig3}
\end{figure*}
\begin{figure*}[t]
\centerline{\includegraphics[width=0.8\textwidth, angle=0]{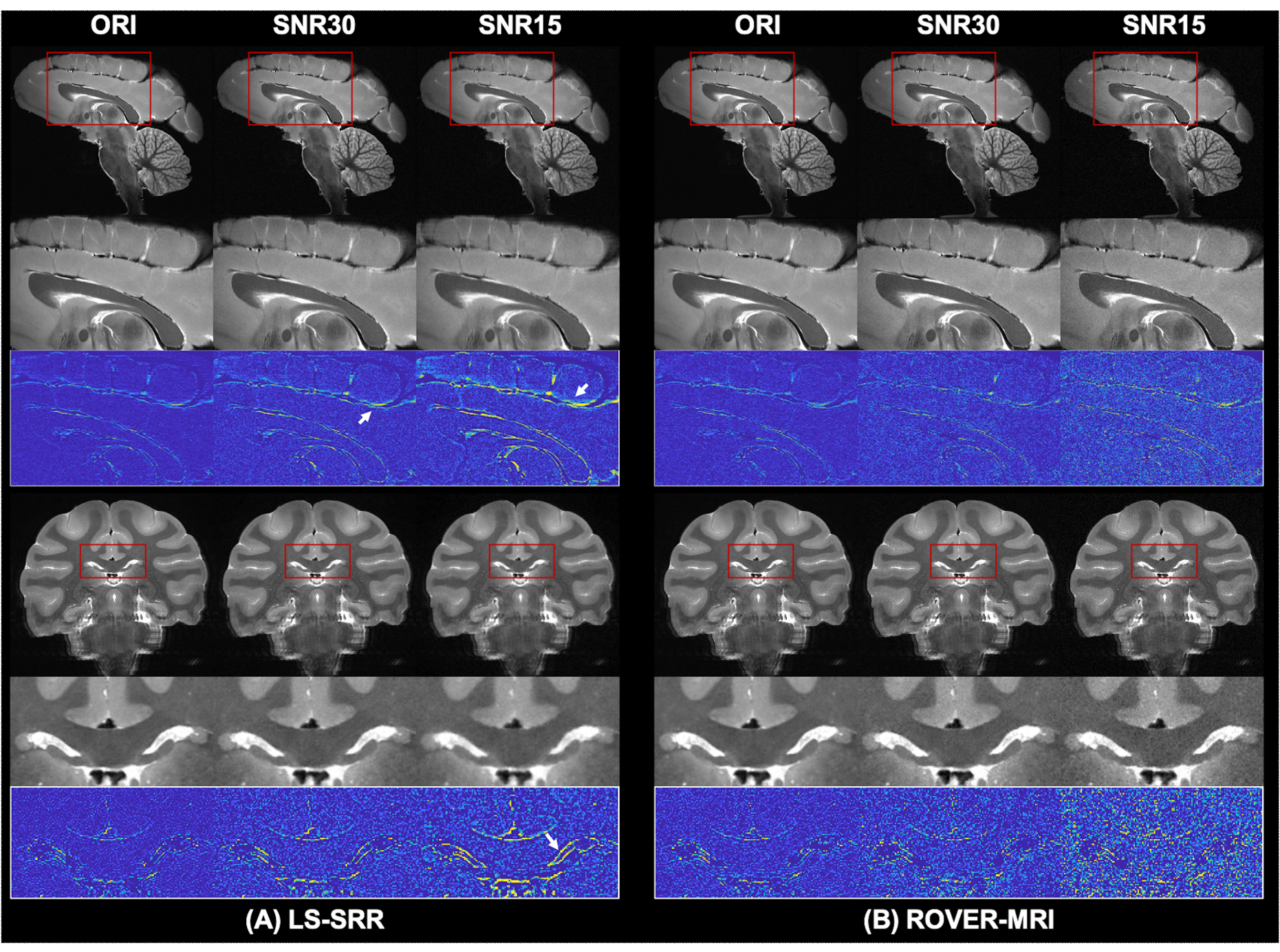}}
\caption{Reconstruction results of our method and LS-SRR with varying noise levels. (A) LS-SRR results under added noise, where reconstructions at SNR 30 show reduced error compared to SNR 15. (B) ROVER-MRI reconstructs cleaner images with significantly reduced errors compared to LS-SRR at both SNR levels. Rows 1 and 4 display typical reconstruction results, while Rows 2 and 5 show enlarged views of the regions within the red boxes. Rows 3 and 6 show the error maps for these regions, emphasizing the robustness of ROVER-MRI against noise.}\label{fig4}
\end{figure*}
\begin{figure}[t]
\centerline{\includegraphics[width=0.5\textwidth, angle=0]{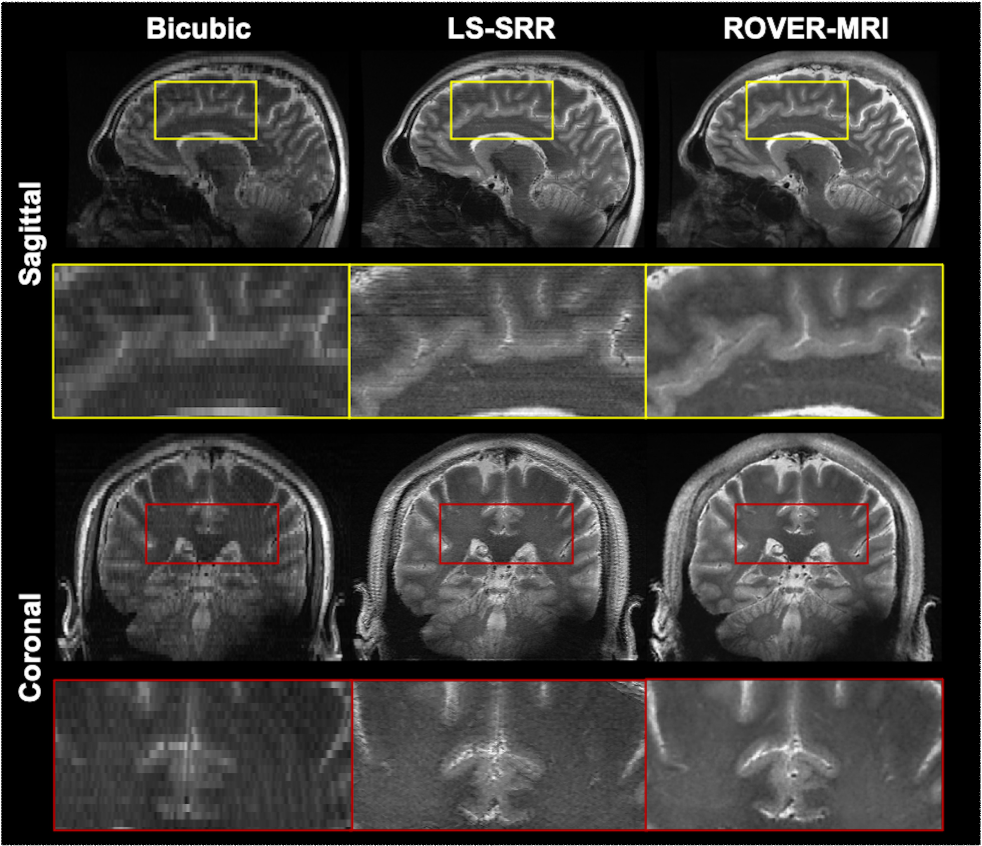}}
\caption{SRR results at 180 µm isotropic resolution. LS-SRR introduces streaking artifacts, which degrade the image quality. In contrast, our ROVER-MRI method demonstrates superior performance, preserving sharper and more continuous anatomical structures while achieving improved SNR. }\label{fig5}
\end{figure}
Next, we evaluated the performance of all methods when fewer than $N_R$ views are used to reconstruct the data. As shown in Figure~\ref{fig3}(A), the LS-SRR method exhibits noticeable boundary artifacts (yellow arrows), and smoothing of fine anatomical details (white arrow) when 6 or 4 views are used. In contrast, Figure~\ref{fig3}(b) demonstrates the performance of ROVER-MRI, which effectively suppresses boundary artifacts and minimizes residual errors, ensuring superior reconstruction quality even with fewer views. Notably, when the number of views is halved to 4, ROVER-MRI maintains high-quality reconstruction, demonstrating its robustness in balancing speed and accuracy. This capability underscores its potential to significantly reduce acquisition time by 2-fold while preserving fine anatomical details.

In Figure~\ref{fig4}, we show the performance in the presence of different levels of noise. The residual maps for LS-SRR reveal prominent errors, as highlighted by the yellow arrows, particularly in regions with fine structural details. These errors become more severe at a lower SNR of 15, indicating the method's vulnerability to noise. Similarly, as shown by the white arrows, residual artifacts persist across the reconstruction, further compromising image fidelity. In contrast, Figure~\ref{fig4}(b) shows the residual maps for ROVER-MRI, which exhibit significantly reduced errors and better preservation of structural details, even under noisy conditions. At both SNR=15 and SNR=30, our method consistently outperforms LS-SRR. 

\begin{table}[t]
\caption{~Hash encoding parameters: hash table size $T$, number of feature dimensions per entry $F$, the coarsest resolution $N_{min}$ and the regularization parameter $\lambda_c$ were set based on the spatial resolution and SNR of the images. }
\label{tab1}
\centering
\begin{tabular}{c|c|c|c|c|c}
\hline
           & L  & T  & F & $N_{min}$                  & $\lambda_c$ \\ \hline
9.4T T2W & 11 & 25 & 2 & $2^{18}$ & 0      \\ \hline
7T T2W P1  & 11 & 25 & 2 & $2^{18}$ & 0      \\ \hline
7T GRE P2  & 11 & 25 & 3 & $2^{16}$ & $2\times10^{-5}$   \\ \hline
3T T2W     & 11 & 24 & 2 & $2^{15}$ & $3\times10^{-5}$   \\ \hline
7T GRE P3  & 11 & 25 & 2 & $2^{18}$ & $0$   \\ \hline
\end{tabular}
\end{table}

\begin{table*}[t]
\caption{~Relative Error (RE) values reflect the errors between the low-resolution ground truth (GT) images and the low-resolution images obtained by back-projecting the reconstructed images from different methods.}
\label{tab2}
\centering
\begin{tabular}{c|cccc}
\hline
RE & 9.4T T2W & 7T T2W P1 & 7T GRE P2 & 3T T2W        \\ \hline
LS-SRR         &0.4515       &   0.6727       &    0.2923     & 0.4313 \\
ROVER-MRI   & 0.3504          &   0.5956        & 0.2601          & 0.4275 \\ \hline
\end{tabular}
\end{table*}

\subsection{Experiment 3: Performance on 7T MRI}
\subsubsection{7T T2w-SPACE MRI with 15 views}
\begin{figure}[t]
\centerline{\includegraphics[width=0.5\textwidth, angle=0]{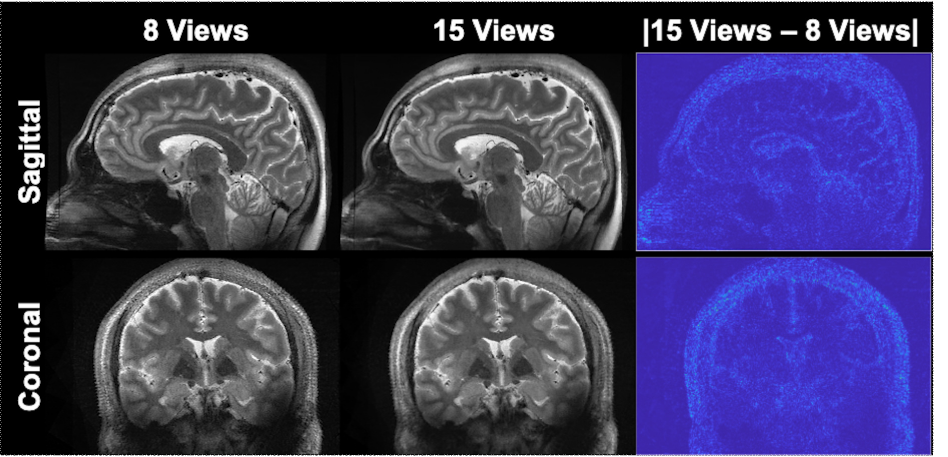}}
\caption{ROVER-MRI results with reduced number of views. The image reconstructed by 15 views was used as a reference. Difference map in the last column shows very little loss in structural details.}\label{fig6}
\end{figure}
Figure~\ref{fig5} illustrates reconstructions at an isotropic resolution of 180 µm, recovering data from a super-resolution factor of 11 in the slice direction. The original LR images exhibit prominent block artifacts, indicative of the resolution limitations inherent to thick-slice acquisitions. While the LS-SRR method improves upon the LR images by enhancing spatial resolution, it introduces pronounced streaking artifacts, which compromises the clarity of fine structures. Additionally, LS-SRR fails to effectively restore fine details, resulting in blurred and discontinuous structures. 
In contrast, ROVER-MRI achieves reconstructions with sharp structural details and enhanced SNR, preserving the integrity of fine anatomical features. According to Table~\ref{tab2}, ROVER-MRI achieves a lower RE, with a reduction of approximately 11.5\% compared to LS-SRR. The method demonstrates its ability to minimize artifacts and recover mesoscale structures with high fidelity, even under challenging super-resolution settings. This underscores the robustness and reliability of ROVER-MRI in producing high-quality images that are critical for accurate interpretation and analysis.

Furthermore, Figure\ref{fig6} presents reconstruction results from datasets with fewer views (8 interleaved views, requiring only approximately 17 minutes of scan time). Remarkably, these results display similar mesoscale details compared to reconstructions from 15 views, highlighting the efficiency of ROVER-MRI in maintaining high image quality with reduced acquisition time. Thus, ROVER-MRI can achieve substantial time savings without sacrificing structural detail, making it a highly practical solution for research applications.

\subsubsection{7T GRE MRI with 9 views}
\begin{figure}[t]
\centerline{\includegraphics[width=0.5\textwidth, angle=0]{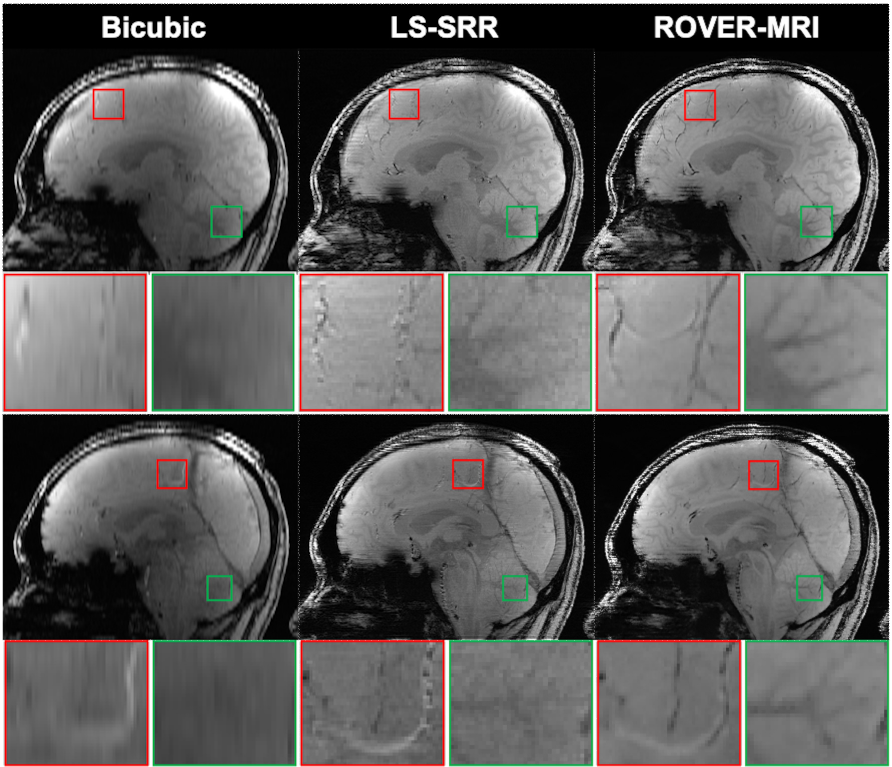}}
\caption{Sagittal brain 7T T2 MRIs reconstructed using three methods: Bicubic, LS-SRR, and ROVER-MRI. The first and third rows display the reconstructed images for each method.
The second and fourth rows present zoomed-in regions (indicated by red and green boxes) for detailed comparison.}\label{fig7}
\end{figure}

Figure~\ref{fig7} illustrates a comparison on 7T GRE data using the three methods. The zoomed-in regions, marked by red and green boxes, highlight key differences in detail preservation and artifact suppression. 
Here too, ROVER-MRI shows superior performance, displaying continuous and artifact-free structures with enhanced sharpness. Table~\ref{tab2} shows that ROVER-MRI achieves a noticeably lower RE with a reduction of approximately 11\%.

\subsubsection{7T GRE MRI with 5 views versus 3D-GRE acquisition}
\begin{figure}[t]
\centerline{\includegraphics[width=0.5\textwidth, angle=0]{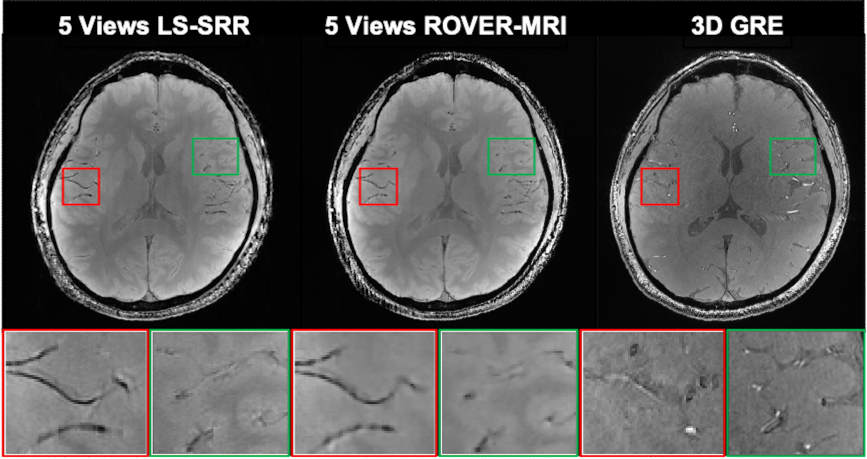}}
\caption{Comparison of reconstruction results from 5 views with 3D GRE acquisition. Both SRR algorithms exhibit finer anatomical details than 3D GRE.}\label{fig8}
\end{figure}
We also compared ROVER-MRI reconstructions from 5 thick-slice views with the data obtained using a 3D-GRE sequence in the same scan time of about 18 minutes. Both LS-SRR and ROVER-MRI outperformed the 3D-GRE data in detail restoration. ROVER-MRI showed superior reconstruction quality compared to LS-SRR. In the red box, ROVER-MRI shows better continuity of blood vessels, and in the green box, it shows a more accurate depiction of anatomical structures. We also note that the SNR is much higher for the ROVER-MRI reconstruction compared to 3D-GRE (although the 3D-GRE scan could be optimized for better SNR by changing the flip angle and relaxation time at the expense of scan time). However, in this study we set the parameters to match the scan time. This experiment demonstrates the utility of ROVER-MRI compared to a full 3D acquisition.

\subsection{Experiment 4: Performance on T2w MRI from 3T scanner}
\begin{figure}[t]
\centerline{\includegraphics[width=0.5\textwidth, angle=0]{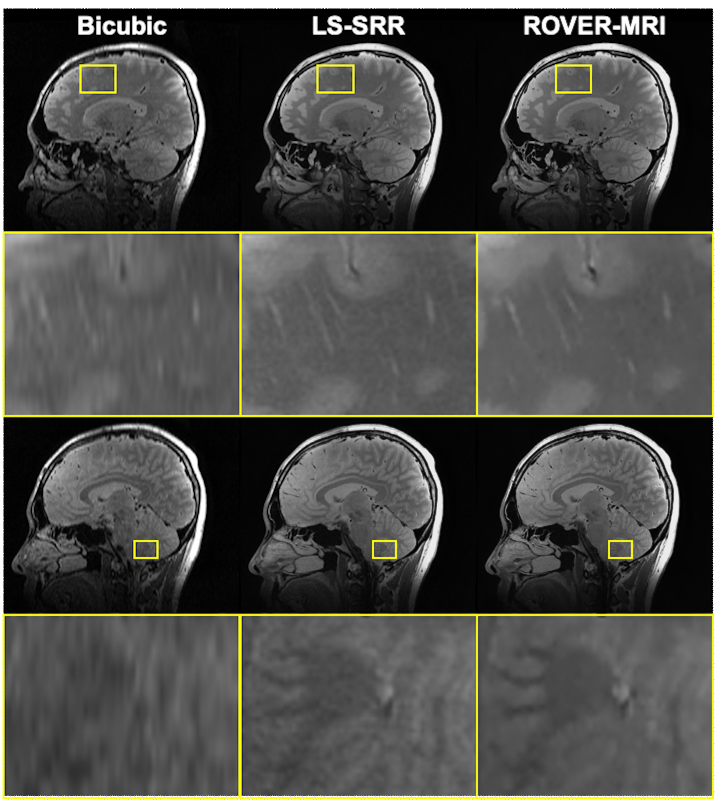}}
\caption{3T T2-weighted brain MRI reconstructed using three different methods. The first and third rows show the reconstruction, while the second and fourth rows present zoomed-in views of the regions marked by yellow boxes. }\label{fig9}
\end{figure}
%
Figure~\ref{fig9} illustrates a qualitative comparison of reconstructed sagittal brain MR images using Bicubic interpolation, LS-SRR, and ROVER-MRI. The zoomed-in views (second and fourth rows) highlight a marked improvement in image quality achieved by ROVER-MRI, with enhanced clarity of fine structures and significantly reduced blurring artifacts. 

\begin{figure}[t]
\centerline{\includegraphics[width=0.5\textwidth, angle=0]{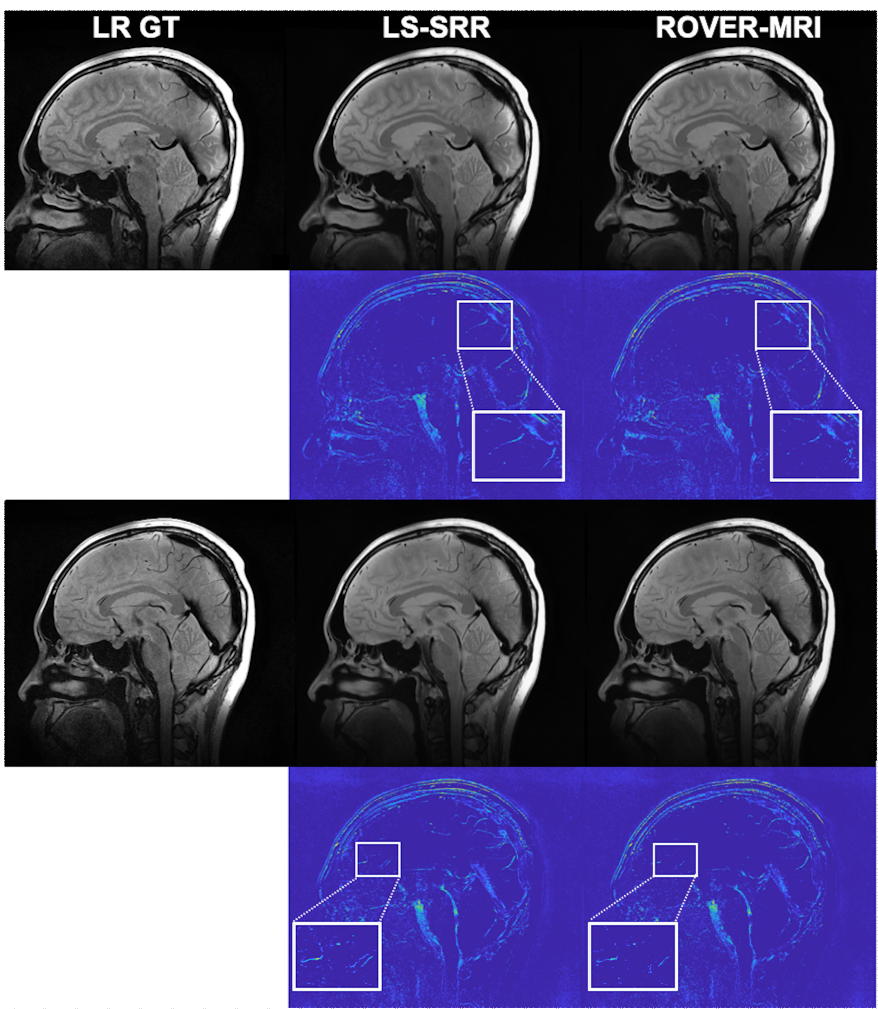}}
\caption{Sagittal brain MR images back-projected into the low-resolution space. The first and third rows display the back-projected images for each method. The second and fourth rows show the error maps calculated against the low-resolution ground truth (LR GT). Zoomed-in regions (indicated by white boxes) highlight structural details for closer inspection. }\label{fig10}
\end{figure}
%
Figure~\ref{fig10} compares sagittal brain MR images back-projected into the low-resolution space. The error maps, along with magnified regions indicated by white boxes, highlight that ROVER-MRI achieves a notable reduction in reconstruction error. These results illustrate that ROVER-MRI not only preserves structural integrity more effectively but also achieves a closer approximation to the low-resolution acquired data, affirming its advantage in reconstruction accuracy.


\begin{figure}[t]
\centerline{\includegraphics[width=0.5\textwidth, angle=0]{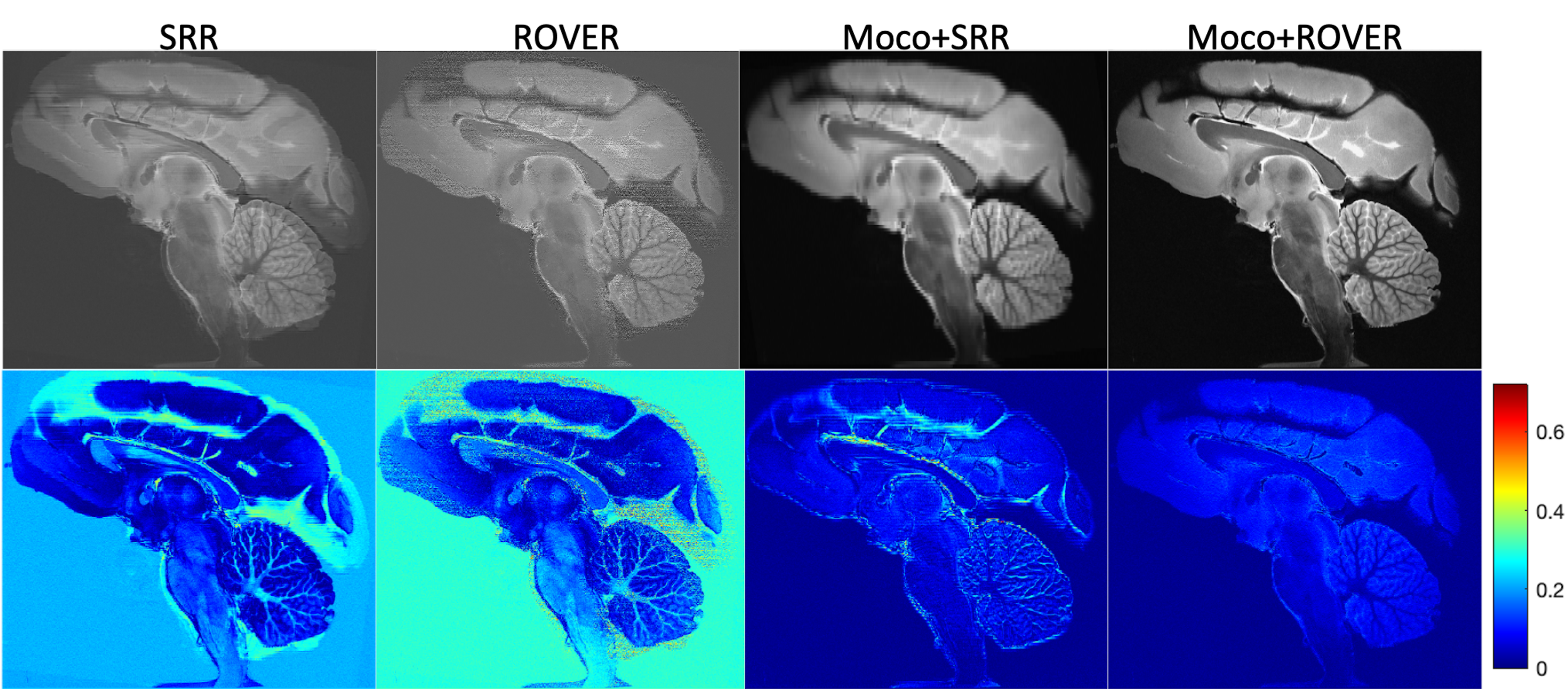}}
\caption{Reconstruction results of simulated monkey brain MRI data with motion (rotations of 5°, 7°, and 10° applied to three of eight views). Results without motion correction (w/o Moco), with motion-corrected LS-SRR (Moco+SRR), and with the proposed motion-corrected ROVER-MRI (Moco+ROVER) are compared against ground truth (GT). The top row displays reconstructed intensity images, and the bottom row shows corresponding absolute error maps (brighter regions indicate higher errors).}\label{moco_mk}
\end{figure}

\begin{figure}[t]
\centerline{\includegraphics[width=0.5\textwidth, angle=0]{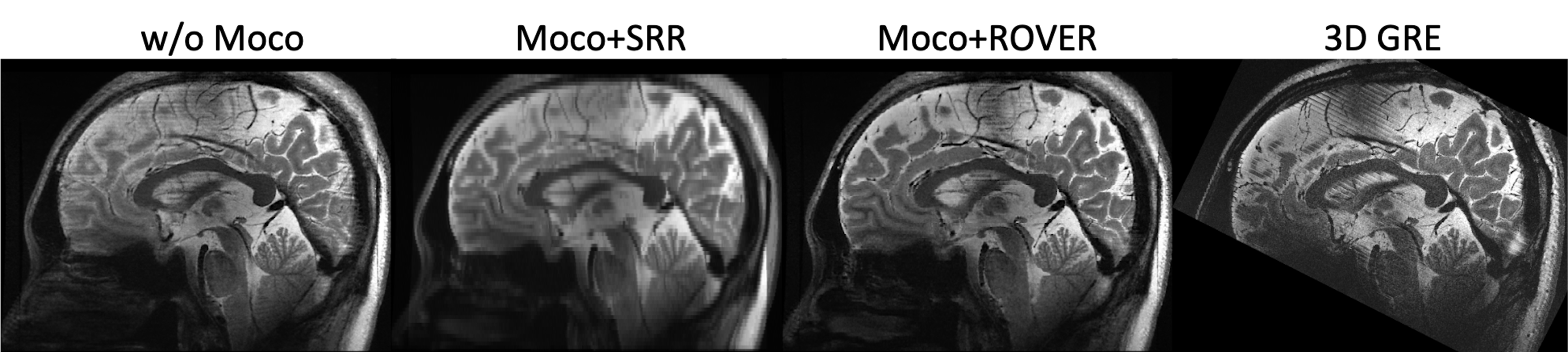}}
\caption{Comparison of in-vivo human brain reconstruction results acquired at 7T MRI without motion correction (w/o Moco), with motion-corrected LS-SRR (Moco+SRR), and motion-corrected ROVER-MRI reconstruction (Moco+ROVER), benchmarked against a high-quality reference 3D gradient echo (3D GRE) image. }\label{moco_hm}
\end{figure}

\subsection{Experiment 5: Performance on Motion-Corrupted Data}
\subsubsection{Simulation experiments with monkey brain data}
Figure~\ref{moco_mk} shows that the proposed motion-corrected ROVER-MRI reconstruction (Moco+ROVER) effectively reduces motion artifacts introduced by simulated rotations (5°, 7°, and 10° applied to three of eight views). Compared to reconstruction without motion correction (w/o Moco) and regularized least-squares super-resolution with motion correction (Moco+SRR), the Moco+ROVER method significantly reduces reconstruction errors and enhances structural accuracy, as clearly shown in intensity images and absolute error maps.
\subsubsection{In-vivo human brain imaging experiments}
Figure~\ref{moco_hm} further validates the performance of Moco+ROVER in realistic clinical settings. Moco+ROVER achieves superior anatomical clarity and structural fidelity compared to methods without motion correction and Moco+SRR reconstruction. In particular, Moco+ROVER shows finer anatomical details compared to the 3D gradient echo (3D GRE) scan, highlighting its effectiveness for rapid, motion-robust high-resolution neuroimaging applications.

\section{Discussion}\label{dis}
In this work, we demonstrated our ROVER-MRI framework on four mesoscale MRI datasets. Among these, the highest resolution achieved was 180 µm isotropic voxel size with high-fidelity results obtained from just 17 minutes of scan time (2x acceleration). Our results highlight significant improvements in spatial resolution, SNR, and acquisition efficiency, addressing longstanding challenges in mesoscale MRI.

The primary advantage of ROVER-MRI lies in its ability to significantly reduce acquisition time while preserving detailed anatomical structures. This is achieved through a combination of multi-view thick-slice acquisitions and an enhanced super-resolution reconstruction framework leveraging multi-resolution hash encoding. Compared to conventional SRR methods such as LS-SRR and bicubic interpolation, ROVER-MRI demonstrated superior reconstruction accuracy, offering sharper structural details and minimal artifacts (Figure~\ref{fig2}$\sim$~Figure~\ref{fig6}). Importantly, our method does not require high-resolution ground-truth datasets for training, providing a flexible and practical solution for high-resolution MRI across diverse imaging protocols. Further, incorporation of motion correction enables robustness to motion during acquisition making the method directly applicable to upcoming studies.

Our experiments further reveal the robustness of ROVER-MRI in scenarios with reduced view counts and low-SNR data. As shown in Figure~\ref{fig3} and Figure~\ref{fig6}, even with 2 times fewer views (scans), reconstruction quality remained high, enabling faster scan times with minimal loss in resolution or accuracy. Additionally, ROVER-MRI outperformed LS-SRR in suppressing noise across varying SNR levels, underscoring its resilience against signal degradation. This robustness is particularly beneficial for clinical and research applications where time constraints or physiological noise might impact image quality.


Despite these advancements, several limitations should be acknowledged. While ROVER-MRI effectively addresses motion artifacts through integrated motion estimation and correction, the reliance on thick-slice acquisitions may still introduce other potential challenges, such as eddy current-induced distortions or residual inaccuracies from imperfect registration. These effects could result in subtle blurring or structural inaccuracies in the final reconstruction, as discussed in prior studies~\citep{shilling2008super, plenge2012super}. Future work could incorporate advanced correction methods for eddy current effects~\citep{szczepankiewicz2019tensor} to further enhance reconstruction fidelity and reliability.

Another limitation pertains to regularization. In this work, we manually fixed the regularization parameter. Although data-driven Bayesian techniques can be used to determine the optimal regularization weight, they can be computationally impractical when dealing with such large datasets. Further, the computational time required for training INR is a bit longer than that required to estimate the data using LS-SRR. We however note that, the computational time for INRs can be significantly reduced using transfer learning techniques. 


In summary, our results demonstrate that ROVER-MRI enables rapid, high-SNR whole-brain imaging at isotropic resolution, offering a significant advancement over existing mesoscale MR imaging methods, especially for T2w images. By integrating multi-view acquisitions with multi-resolution hash encoding, our approach achieves high-quality super-resolution reconstruction while remaining robust to noise and motion. These capabilities make ROVER-MRI well-suited for mesoscale neuroimaging studies and hold great potential for applications requiring time-efficient, high-resolution whole-brain imaging.

\section*{Acknowledgments}
This study is supported by National Institutes of Health grants: R01 NS125307, R01MH132610, R01MH125860, R01EB032378,  R01MH116173 and R01MH119222.

\bibliographystyle{model2-names.bst}\biboptions{authoryear}
\bibliography{refs}

\end{document}